\documentclass{article}
\usepackage{amsmath}
\usepackage{hyperref}
\usepackage{epsfig}
\usepackage{graphicx,caption}
\usepackage{comment}
\usepackage{bm}
\def\dsize{\displaystyle}
\def\E{\mathop{\rm E}\nolimits}
\def\Ro{\mathop{\rm Ro}\nolimits}

\def\Ra{\mathop{\rm Ra}\nolimits}

\def\Pr{\mathop{\rm Pr}\nolimits}
\def\q{\mathop{\rm q}\nolimits}

\begin{document}
\begin{center}{
\bf  LATITUDE DEPENDENCE OF CONVECTION \\ AND MAGNETIC FIELD GENERATION IN THE CUBE}

\end{center}

\begin{center}{M. Yu. Reshetnyak  \\ 
\vskip 0.5cm

 Institute of Physics of the Earth of RAS,
  Moscow, Russia, 
  
     Institute of Terrestrial Magnetism, Ionosphere and Radio Wave Propagation of RAS, Moscow, Russia, 
 m.reshetnyak@gmail.com}
\end{center}

       \abstract{\noindent  The 3D thermal convection in the Boussinesq approximation with heating from below and dynamo in the cube are considered. We study dependence of the  convection intensity and magnetic field generation on the latitude in $\beta$-plane approximation.               
                    It is shown that kinetic energy gradually increases from the poles  to the equator more than order of magnitude. The model predicts the strong azimuthal thermal wind, which direction depends on the sign of the thermal convective fluctuations. The spatial scale of  the arising flow is comparable to the scale of the physical  domain. 
             The magnetic energy  increases as well, however dynamo efficiency, i.e., the ratio of the magnetic energy to the kinetic one decreases to  the equator.
             This effect can explain predominance of the  dipole configuration of the magnetic field observed in the planets and stars.
      The approach is useful for modeling of the magnetohydrodynamic 
        turbulence in planetary cores and stellar convective zones.

      \vskip 0.1cm}

 \section{Introduction} 
   3D modeling of convection and dynamo processes in the planetary cores and convective zones of the stars is a modern branch of the magnetohydrodynamic  (MHD) simulations 
   \cite{RHK2013}. There are two approaches which are usually used. For the large-scale fields modeling the MHD equations are solved in the spherical geometry, see, e.g.,
   \cite{Jones},  Inclusion of the realistic boundary conditions and distributions of the energy sources in the model let to simulate  the various observable features of the magnetic fields, like the spatial and temporal spectra of the magnetic fields, the butterfly diagrams in the case of the solar dynamo, reversals of the geomagnetic field.

              However similarity of observations and simulations sometimes can be misleading because the parameters used in 3D models are still quite far from the desired ones. Briefly, the main problem is a turbulence which simulation requires resolution of the small scales. The other difficulty is the anisotropy of the flow, concerned with the rapid daily rotation, as in the case of the planetary cores. The spherical models which have dense distribution of the mesh grid points near the poles,  require the small time step (because of the Courant condition), and therefore  do not suit to the turbulence modeling. Meanwhile,   simulations in the Cartesian  geometry with  the homogeneous grids could be very helpful. These simulations reproduce the cascade processes between the scales, are helpful for estimates of the turbulent coefficients, and demonstrate  various  remarkable properties of the MHD turbulence  \cite{BS2005}.

            In spite of the fact that these two approaches were used for years, the flat layer dynamo simulations  did not take into account the latitude dependence to the moment. In other words the angle between the angular rotation axis $\bm \Omega$  and gravity $\bf g$  was set to zero 
\cite{JR2000}, \cite{B2003}, 
      \cite{CEW2003}.     
              This choice of parameters  corresponds to the geographic poles. Having in mind that the most common configuration of the magnetic field in planets and stars is the dipole, and  3D spherical dynamo models  predict existence of the   toroidal counterpart,  concentrated  at the mean latitudes, it looks tempting to include the latitude dependence in the flat layer models immediately. The lack of the papers, devoted to  this problem  in the geodynamo and stellar  applications, looks quite surprising because in the meteorology  the latitude dependence, known as the $\beta$-plane approximation, was already used  in the Cartesian models  for a long time
 \cite{P2012}. 

                The other motivation of this paper is to distinguish the physical effects related  to the angle between $\bm \Omega$ and $\bf g$ definitely,  leaving apart influence of the inner core and spherical boundaries, which can change hydrodynamics\footnote{The sign of  the boundaries curvature defines the direction of the Rossby waves propagation 
     \cite{B2002}.}, and the magnetic field generation  substantially.

     Below we consider the standard  3D MHD model  in the rapidly rotating cube. The model includes the thermal convection equations  in the Boussinesq approximation with the  heating from below. The liquid is conductive and at the large magnetic Reynolds numbers the fluid motions can generate the magnetic field.  To solve these equations we use MPI C++ code, based on the predictor-corrector method  
    \cite{CHZ1988}. We check how these processes depend on the angle between  $\bm \Omega$ and $\bf g$. Some analytical predictions, based on the analogy with  the motion of the charged particle in the electromagnetic field  are considered as well.

    \section{Dynamo in the cube and  numerical methods}  

The dimensionless  dynamo equations for an incompressible fluid 
  ($\nabla\cdot{\bf V}=0$) in the cube    of  the  height  $\rm L=2\pi$,
  rotating with the angular velocity $\bm \Omega$,  in the Cartesian system of
  coordinates $(x,\,y,\,z)$
   have the form:
   \begin{equation}\begin{array}{l}\dsize
{\partial {\bf A}\over\partial t}={\bf V}\times {\bf B}
+ {\rm q}^{-1}\Delta {\bf A}, \qquad {\bf B}={\rm rot\, }{\bf A}
\\  \\ \dsize
\E\Pr^{-1}\left[ {\partial {\bf V}\over\partial t}+
  \left({\bf V} \cdot \nabla\right){\bf V} \right]
  =  \\  \\ \dsize
 {\rm rot\,}{\bf B}\times {\bf B}
   -\nabla { P} -{\bf
    1_\Omega}\times{\bf V} + \Ra { T} \,{\bf{1}_z}+ \E\Delta {\bf V}
\\ \\ \dsize
{\partial { T}\over\partial t}+\left({\bf V}\cdot\nabla\right)
\left({ T}+{ T}_0\right)= \Delta { T}.
\end{array}\label{sys00}
\end{equation}

 Velocity $\bf V$, magnetic field $\bf B$ (derived  from the vector potential $\bf A$), pressure
$P$ and the typical
diffusion time $t$ are measured in units of
$ \rm \kappa/L$,   $\dsize \sqrt{2\Omega\kappa\mu\rho};$, $\rm
\rho\kappa^2/L^2$ and $\rm L^2/\kappa$ respectively, where $\kappa$ is
the thermal  diffusivity, $\rho$ is the density, $\mu$ the permeability,
$\dsize \Pr=\frac{\kappa}{
\nu}$ is the Prandtl number,
   $ \dsize \E =\rm  \frac{\nu}{ 2\Omega L^2}$ is the
Ekman number, $\nu$ is the kinematic viscosity,
$\eta$ is the magnetic diffusivity, and ${\q}=\kappa/\eta$ is the
Roberts number.
$\dsize \Ra\rm
 =\frac{\alpha g\delta T {L}}{  2\Omega\kappa}$ is the modified
Rayleigh number, $\alpha$ is the coefficient of the volume expansion,
  $\delta T$ is the unit of the temperature fluctuations $T$,  $g$ is the gravitational acceleration, and $T_0=2\pi-z$ is the temperature profile, corresponding to the heating  from below.
  
   The unit vector $\bf 1_{\bm \Omega}$ defines direction of the angular velocity $\bm \Omega$.
   The angle between   $\bm \Omega$  and gravity $\bf g$,  directed along the $z$-axis, is equal to the colatitude $\theta$, which is related to  the latitude as $\vartheta=90^\circ-\theta$. 

  The problem is closed with the periodical boundary conditions in the $(x,\, y)$-plane. In  $z$-direction the  following simplified boundary conditions
\begin{equation}
\begin{array}{l}\dsize
\dsize T= V_z=A_z={\partial V_x\over \partial z}=
  {\partial V_y\over \partial z}={\partial A_x\over \partial z}={\partial A_y\over \partial z}=0
\end{array}\label{sys1}
\end{equation}
   at $z=0,\,2\pi$ are used. Conditions for $\bf A$ are the so-called pseudo-vacuum boundary conditions, correspond to the following conditions for the magnetic field:   
   $\dsize B_x=B_y={\partial B_z\over \partial z}=0$. Then the  normal component to the boundary of the electric current $ {\bf J}={\nabla}\times{\bf B}$ is zero. 

   The system  \ref{sys00} was solved using the finite differences of the second-order in space and time. 
   For approximation of the time derivative  the three-layer time scheme was used:
  \begin{equation}
\begin{array}{l}\dsize
       {
   \partial f \over \partial t
   }
    ={3f^{n+1}-4f^n+f^{n-1}\over 2\, \delta t},
\end{array}\label{sys11}
\end{equation}
           where  $n$ denotes the time step $\delta t$.

 For  $T$  and $\bf A$ the corresponding equations from  \ref{sys00}  were written in the form:

  \begin{equation}
\begin{array}{l}\dsize
    {3f^{n+1}-4f^n+f^{n-1}\over 2\, \delta t}=2\,{F}^n-{F}^{n-1}+{1\over 2}\nabla^2{f}^{n+1},
\end{array}\label{sys111}
\end{equation}
where 
\begin{equation}
\begin{array}{l}\dsize
     {F}^n= {C}^n+{1\over 2}\nabla^2{f}^{n},
  \end{array}\label{sys101}
\end{equation}
 and $C$ denotes the corresponding convective term.
 Eqs(\ref{sys111},\ref{sys101}) with respect to $f^{n+1}$ were solved using the Gauss-Seidel method.

  While using the vector potential $\bf A$ provides divergence free of the magnetic field $\bf B$, incompressibility of the velocity field $\bf V$ should be provided by some special technique. Here we use  the predictor-corrector  
   method \cite{CHZ1988},  \cite{Bandaru2016}, introducing the intermediate velocity field $\bf V^*$ by equation:
 \begin{equation}
\begin{array}{l}\dsize
        {3\,{\bf V}^*-4\,{\bf V}^n+{\bf V}^{n-1}\over 2 \,\delta t}=2\,{\bf F}^n-{\bf F}^{n-1}+{1\over 2}\nabla^2{\bf V}^*, 
\end{array}\label{sys11l}
\end{equation}
where 
\begin{equation}
\begin{array}{l}\dsize
       {\bf F}^n= 
-\E\Pr^{-1}      \left({\bf V}^n \cdot \nabla\right){\bf V}^n   
+
 {\rm rot\,}{\bf B}^n\times {\bf B}^n-
       {\bf
      1_\Omega}\times{\bf V}^n + \Ra { T}^n \,{\bf{1}_z}+{1\over 2} \E\Delta {\bf V}^n.
 \end{array}\label{sys10l}
\end{equation}
Eqs\ref{sys11l},\ref{sys10l} lead to a Poisson-type equation for ${\bf V}^*$.

 Then   pressure  $P^{n + 1}$ was  derived  from the
 continuity equation by solving the another Poisson problem:
\begin{equation}
\begin{array}{l}\dsize
    \nabla^2 P^{n+1}={3\over 2\, \delta t}\nabla\cdot {\bf V}^*
 \end{array}\label{sys12}
\end{equation}
 with the Neumann boundary condition for $z$-coordinate:
\begin{equation}
\begin{array}{l}\dsize
    {\partial P^{n+1}\over \partial z}=
        {3\over 2\, \delta t}V_z^*.
 \end{array}\label{sys13}
\end{equation}
The last step provides  incompressibility of the velocity field:
\begin{equation}
\begin{array}{l}\dsize
    {\bf V}^{n+1}={\bf V}^{*}- {2\delta t\over 3}\nabla P^{n+1}.
  \end{array}\label{sys14}
\end{equation}

The second order up-wind scheme was used for approximations of the convective terms in the heat and the Navier-Stokes equations:
 \begin{equation}
    V_i{\partial \over \partial x}T_i=
\begin{cases}
         (3T_i-4T_{i-1}+T_{i-2})V_i/(2\delta x), \,\,\,\,\,\, V\ge 0 \\
       (-T_{i+2}+4T_{i+1}-3T_i) V_i/(2\delta x), \,\, V< 0,
   \end{cases}\label{rr}
 \end{equation}
 where $i$ denotes the index of the grid step $\delta x$. The scheme \ref{rr} was used for $y$-,$z$-directions in the same way.

               This approach was realised in C++ code with MPI. The whole domain was divided into $(N\times N)$,  subdomains in $(x,\,y)$ coordinates, where the MHD equations \ref{sys00} were solved. The subdomains exchanged by its boundaries at the each time step $n$. In this paper  the mesh grid $(295\times 295\times 125)$ in $(x,\,y,\,z)$ coordinates, and $N=6$ were used. The simulations were done at two linux workstations 
            Intel(R) Xeon(R) CPU E5-2640 with  40 cores and 128Hb common memory  at the station. Each run required 2-4 days, depending on the time step $\delta t$, which was in the range $(2\div 5)\, 10^{-7}$.

\section{Pure convection}

     Before to consider the full dynamo  the pure  turbulent convective  regime with $\E=4\, 10^{-5}$, $\Ra=9\, 10^3$, $\Pr=1$ was studied. Ten runs with step $10^\circ$ in the latitude $\vartheta$ were performed. After some intermediate stage  solutions reached the quasi-stationary states. To measure 
      the intensity of convection we estimated  the mean over the volume kinetic energy, see Figure 1,  defined as  
\begin{equation}
\begin{array}{l}\dsize
      E_k={1\over {2 \cal V} }\int\limits_{\cal V} {\bf V}^2 \, d\, {\bf r}^3,\qquad {\cal V}=8\,\pi^3. 
    \end{array}\label{sys14}
\end{equation}
\begin{figure}[ht!]   
     \def \ss {10cm}   
\captionsetup{width=.8\linewidth}
       \vskip -4cm
  \hskip 3.2cm \epsfig{figure=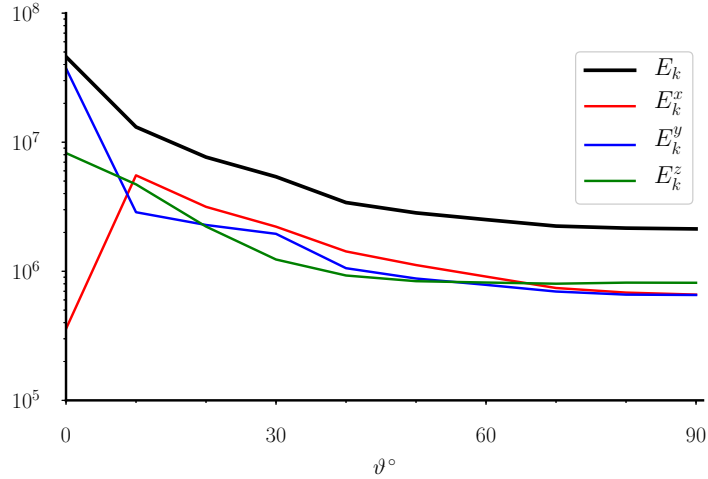,width=\ss}
        \vskip -2.8cm
                     \caption{
                                The latitude dependence of the kinetic energies. $E_k^x$,  $E_k^y$,  $E_k^z$ denote the kinetic energies of $V_x$-, $V_y$-, and $V_z$-components of the velocity, respectively. The total kinetic energy  $E_k=E_k^x+E_k^y+E_k^z$.
                                  } \label{1}
                                  \end{figure}
    
                                  The considered regimes correspond to the developed turbulence with the Reynolds number 
                                   ${\rm Re}=2\pi\, \sqrt{2\, E_k/3}\gg 1$.    
         All the energies increase from the poles to the equator. 
               Only $E_k^x$ at the equator has sharp minimum. This behaviour is expectable because at the equator, $\vartheta=0$, $x$-components of the Coriolis and Archimedean forces are zero, and the non-zero value of $V_x$  is provided by  the non-linear term in the Navier-Stokes equation only. The ratio of the maximum and minimum of  $E_k$   is $22$. This effect is quite strong and should be explained in some way.


\begin{figure}[ht!]   
     \def \ss {9cm}   
\captionsetup{width=.8\linewidth}
       \vskip -3.2cm
  \hskip 0.2cm \epsfig{figure=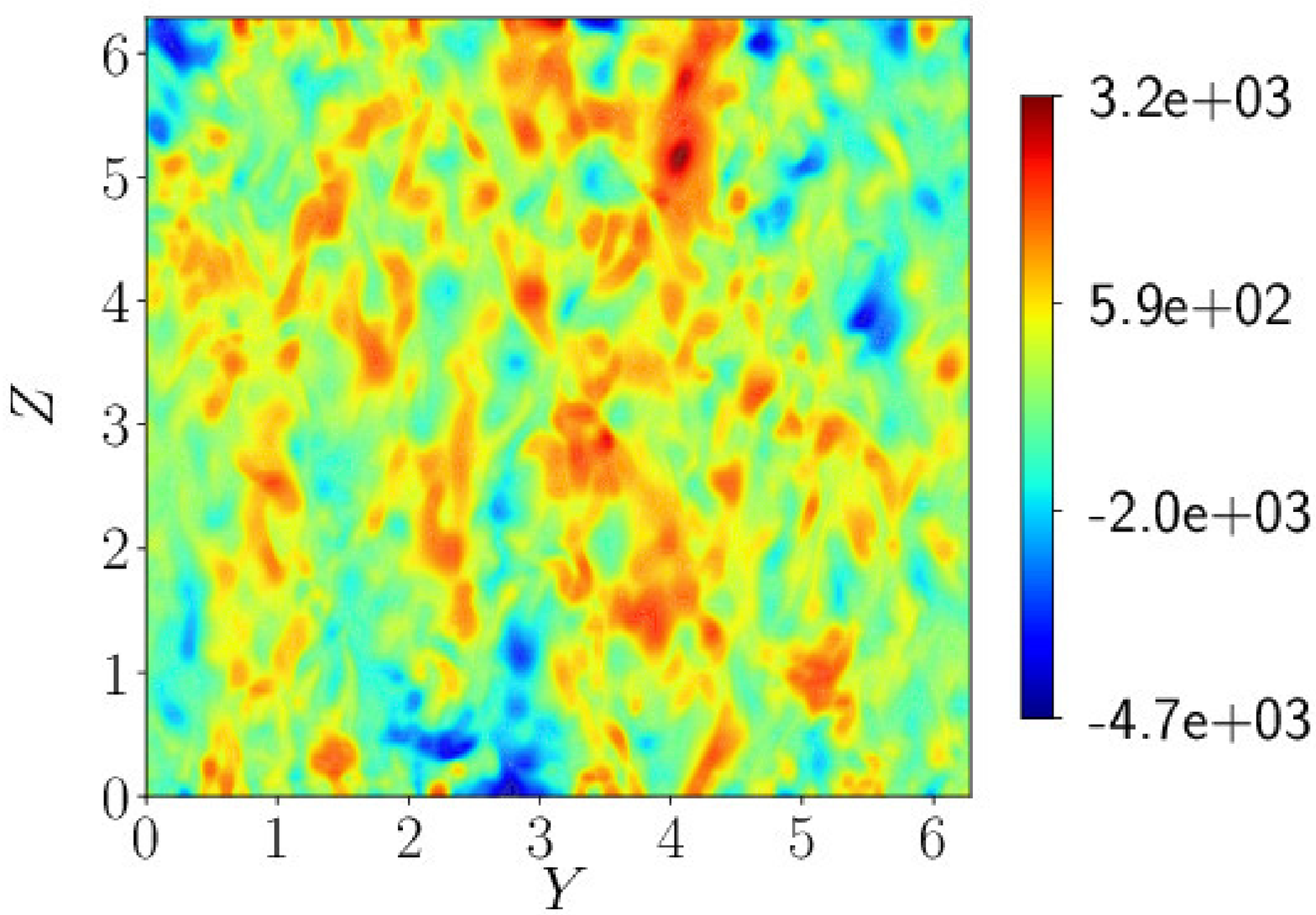,width=\ss}
  \hskip 0.2cm \epsfig{figure=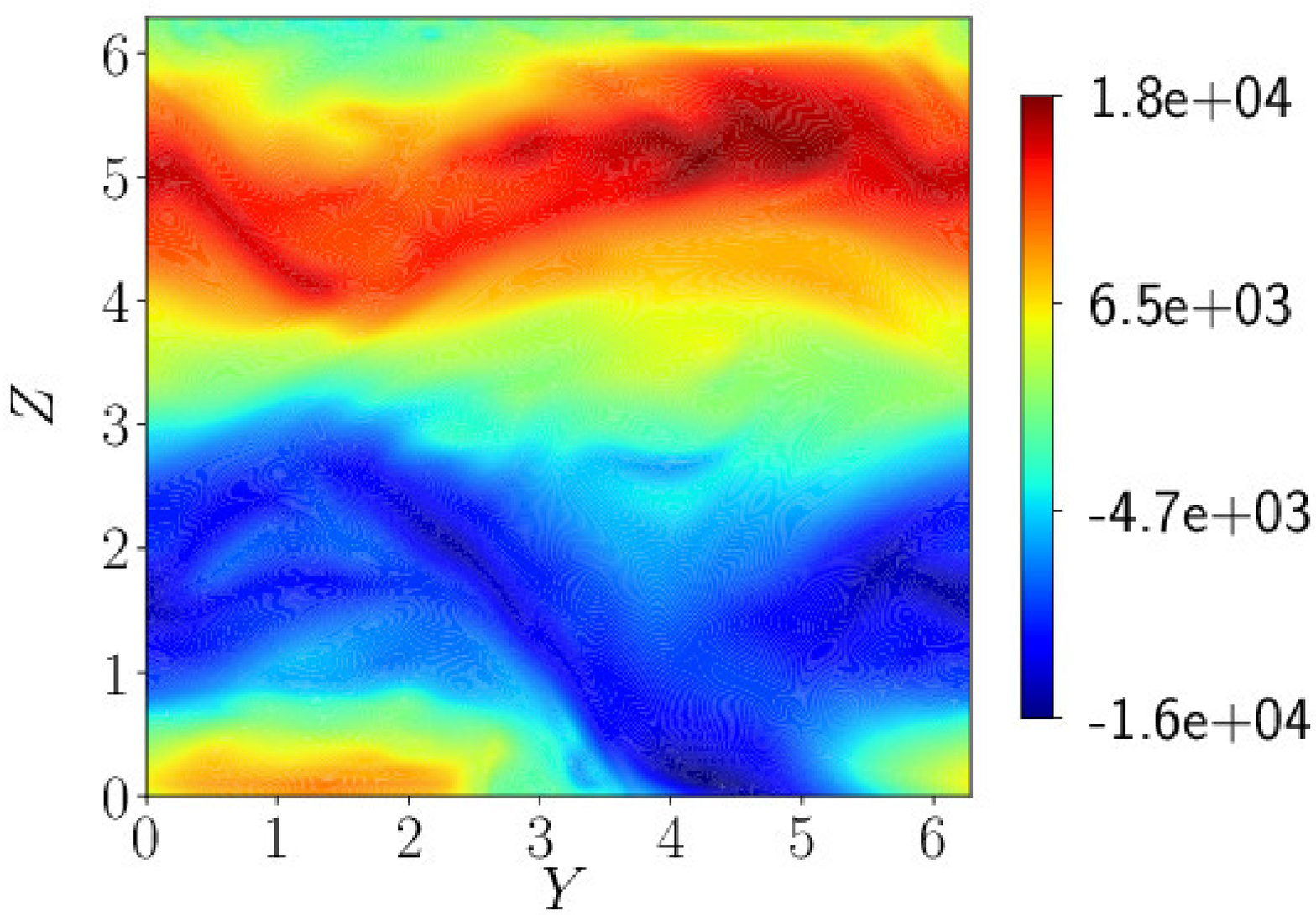,width=\ss}
  \vskip -2.8cm
  \hskip 0.2cm \epsfig{figure=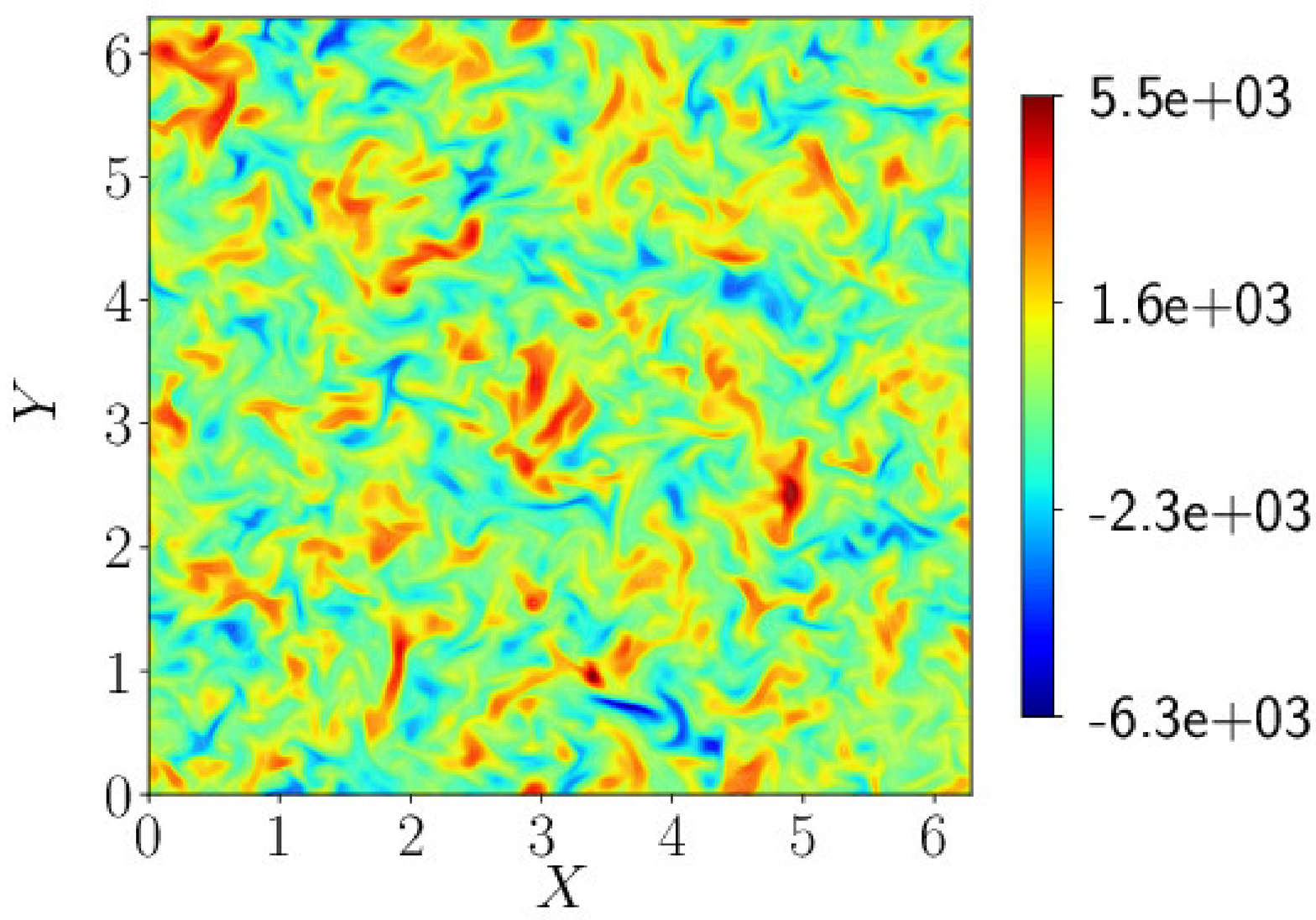,width=\ss}
  \hskip 0.2cm \epsfig{figure=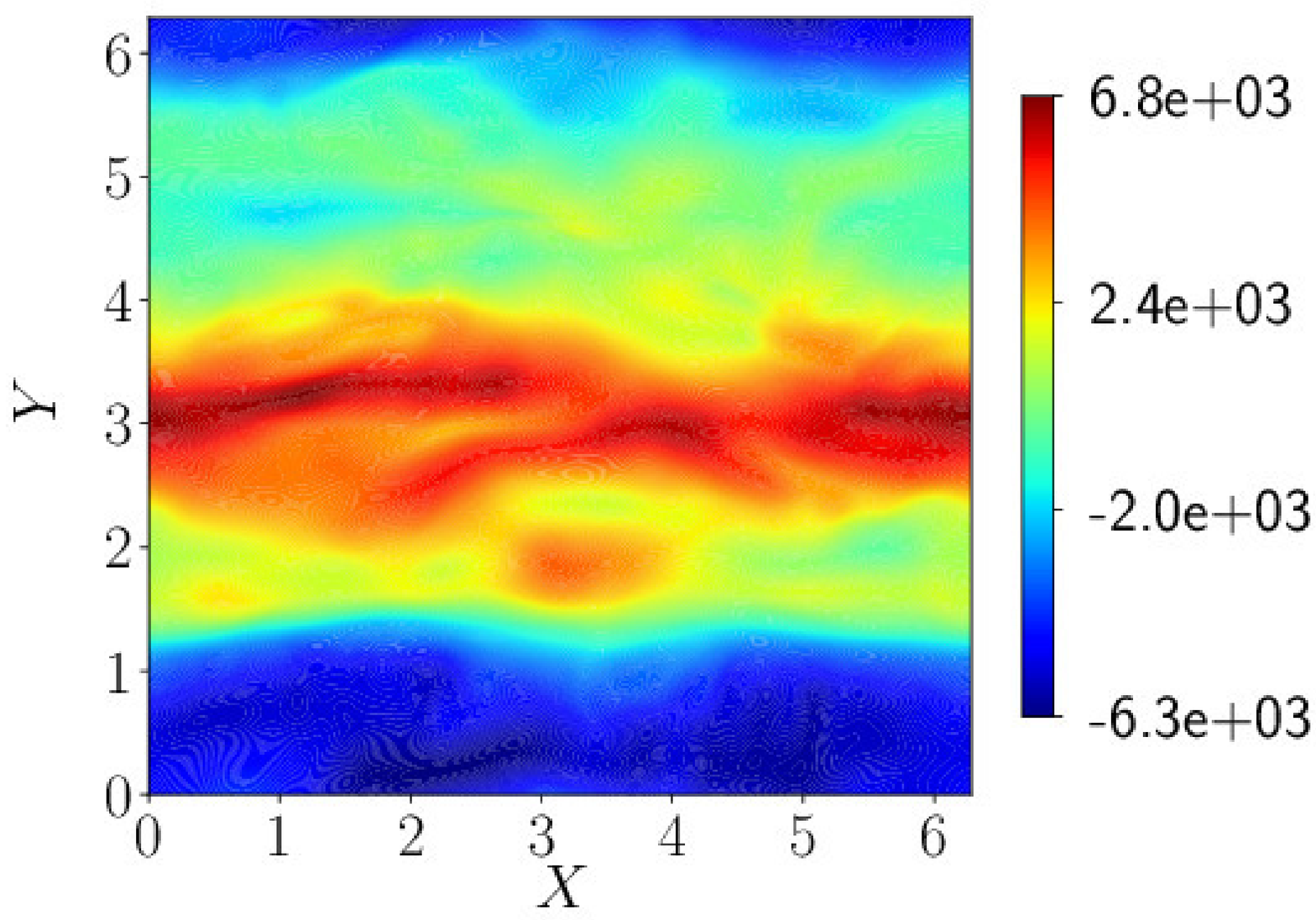,width=\ss}
        \vskip -2.8cm
  \caption{
    The    $x$-sections of $V_y$-components (upper line) and $z$-sections of $V_z$-components (bottom) of the velocity field. The left column corresponds to the pole, and the right one  -- to the equator.
    } \label{2}
 \end{figure}

             Moreover, analysis of the spatial structure of the flows, see Figure 2, reveals that the small-scale cyclonic convection, existed at the poles, changed to the large-scale convection  at the equator. Note, that the Reynolds number in the latter case is larger, and the flow is ``more'' turbulent, but in the same time it is large-scale. 
   It should be noted that $V_y$-component at the equator is perpendicular to $\bm \Omega$ and it should be twisted  at the small scale in the similar way, as it was at the poles.
   
    To find the origin of the large-scale convection the analogy with the motion of the charged particle in the constant electromagnetic field, considered in the next section, is instructive.

\section{Analogy with the charged particle moving in the electromagnetic field}

Similarly to the motion of the charged particle in the constant in time and homogeneous in space 
  electromagnetic field, see  
       \cite{AL1972}, we consider two limiting cases, where Archimedean force and angular rotation velocity of the system are directed along the same axis, and the other, where these forces  are perpendicular. 

   In the first, the  most studied case, which corresponds to the geographic poles, the flow is accelerated by the Archime\-dean force 
     $\sim \Ra T$. Due to the Coriolice force any motion in the orthogonal plane to the gravity and $\bm \Omega$ is twisted with radius $r$, defined by relation  $\Ro V_\perp^2/r\sim V_\perp$, i.e. $r\sim \Ro V_\perp  $, where  $ V_\perp$ is the velocity orthogonal to $\bm \Omega$.
       The Rossby number $\Ro=\Pr^{-1}\E$ for the Earth's core $\sim 10^{-15}$. Even with $V_\perp\sim {\rm Re}= 10^9$, estimated from the large-scale velocity, based on the west drift of the geomagnetic field, one has $r\sim 10^{-6}$ in units of the liquid core's scale. Taking into account decrease of the kinetic energy spectrum will only decrease estimate of $r$. 
     This estimate  is valid for the large velocities with negligible viscous dissipation.  
 
 The linear analysis at the  threshold of convection generation, where  viscous diffusion is important, 
    also predicts existence of the  small scale  in the perpendicular plane: $r\sim \E^{1/3}= 10^{-5}$ 
         \cite{Roberts},   
      \cite{Busse}. 
     
       The both  estimates demonstrate that the small-scale convection at the poles is a quite natural phenomenon, and it appears in the rapidly rotating objects with $\Ro=\E\cdot \Pr\ll 1$     
      even at the critical Rayleigh numbers.

        For the other case, which corresponds to the equator plane, let Archimedean force is still directed along the $z$-axis and $\bm\Omega$ along  $x$, and the  initial velocity is at the  $(y,\,z)$-plane. Then the trajectory of the particle   remains  in the same plane, and its motion is described by equations:
      \begin{equation}\begin{array}{l}\dsize
           \Ro \ddot{y}=\dot{z} \\
           \Ro \ddot{z}=-\dot{y} +\Ra T,
\end{array}\label{sysk0}
\end{equation}
   where dot is for the time derivative.

    Eqs(\ref{sysk0}), written in the reference system moving with the velocity  $v$ along the $y$-axis, after  substitution $y_1=y-vt$, have the form:
   \begin{equation}\begin{array}{l}\dsize
           \Ro \ddot{y_1}=-\dot{z} \\
              \Ro \ddot{z}=\dot{y_1} +v +\Ra T.
 \end{array}\label{sysk}
 \end{equation}
             Choosing  $v=-\Ra T$, one has circular motion in  $(y_1,\,z)$-plane with frequency $\sim \Ro^{-1}$. The trajectory in the original  $(y,\,z)$-plane is a trochoid, i.e. the superposition of the circular motion and the drift with velocity $v$. In terms of the spherical geometry $y$ corresponds to the azimuthal direction.  Addition of the initial velocity in $x$-direction, which  remains constant because of absence of forces in this direction, does not change situation.
       
              In  (\ref{sysk}) $T$ is a fluctuation of the temperature relative to the non-convective distribution, and it can change the sign. By analogy with the motion of the charge in the magnetic field one has thermal separator, which divides the hot ($T>0$, $v<0$)  and cold ($T<0$, $v>0$) flows.
      
            Note that estimates of the radius of rotation in planes perpendicular to the axis of rotation coincide in the both cases. The difference  is existence of a thermal wind with a velocity $ v $ in the azimuthal direction in the latter case. It is this wind has large spatial scale, already detected in Figure 2 at the equator. Due to the continuity equation the other components of the velocity  can also posses the large-scale counterpart, as we observe it  in  $V_z$-flow.
 
\section{Dynamo}

    Starting from the pure convection velocity and temperature distributions, obtained in Section 3,  and the small  magnetic field initial seed,  the full dynamo system \ref{sys00} were integrated in time up to the state where all the physical fields stabilised at the quasi-stationary regime. The corresponding estimates of the magnetic energy, defined as
\begin{equation}
\begin{array}{l}\dsize
     E_m={1\over {2\Ro \cal V} }\int\limits_{\cal V} {\bf B}^2 \, d\, {\bf r}^3,
      \end{array}\label{sys14m}
\end{equation}
 are ploted versus lattitude $\vartheta$ in Figure 3.

\begin{figure}[ht!]   
     \def \ss {10cm}   
\captionsetup{width=.8\linewidth}
       \vskip -5.8cm
  \hskip 2.2cm \epsfig{figure=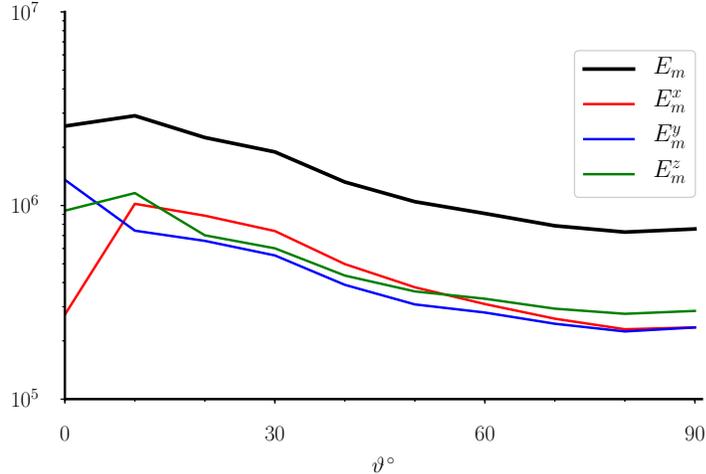,width=\ss}
                   \vskip -2.3cm
                     \caption{
                            The latitude dependence of the magnetic energies. 
                                 } \label{3}
                                  \end{figure}
      The behaviour of the magnetic energy  is similar to  the kinetic energy up to some details near the equator plane. The largest increase of the magnetic energy with the latitude  demonstrates $B_y$-component, which is stretched by the strong large-scale thermal wind $V_y$.
    In contrast to the total kinetic energy, $E_k$, the magnetic energy $E_m$ slightly decreases at the equator. In some sense it resembles behaviour of the magnetic field in the spherical models with the odd configurations of the magnetic field with respect to the equator, e.g., dipole. 
     Increase of the magnetic energy $E_m$  relative to the poles can reach factor 4.
\begin{figure}[ht!]   
     \def \ss {10cm}   
\captionsetup{width=.8\linewidth}
       \vskip -2.8cm
  \hskip 2.2cm \epsfig{figure=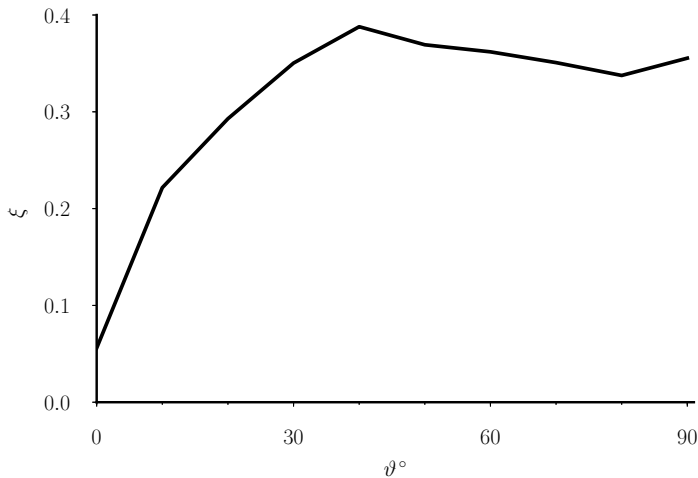,width=\ss}
                    \vskip -2.3cm
                     \caption{
  The  latitude dependence of   the magnetic field generation efficiency $\xi$. 
                                 } \label{4}
                                  \end{figure}

                                    One can expect that the large-scale flow, based on the thermal wind, and the small-scale cyclonic convection have
     different efficiencies of the magnetic field generation. To test this hypothesis the ratio $\xi=\dsize {E_m\over E_k}$ was plotted as a function of $\vartheta$ in Figure 4. It appears that in the range of  $50^\circ<\vartheta<90^\circ$ 
  $\xi$ is approximately constant and then decreases to the equator in one order of the  magnitude. We conclude that the large-scale flow with the small gradients is less  effective than the cyclonic convection with the non-zero net helicity  \cite{R2017}.

  \section{Conclusions} 

         Our simulations clearly demonstrate that the angle between the axis of rotation and gravity changes not only the magnitude of the mean parameters in the flat model, like energies, but the structure of the flow, its spectra, as well. These effects are already notable at the mean latitudes, where the magnetic energy maximum is localised in the spherical models, and should be taken into account in future. Of course, due to variety of the physical effects, application of these results to  the spherical shells, should be done carefully.  Thus existence of the well-known in geodynamo  since  
       \cite{GRPEPI95}, see also 
         \cite{RP2016},      
     the  large-scaled vortexes in the Taylor cylinder at the large Rayleigh numbers,   introduce the new complexity in the model.

     The other issue is improvement of the numerical methods for the MHD turbulence modling. 
      The finite difference methods are the most promising approach for the multi-core simulations. The modern higher order approximations of MHD differential equations are already comparable by accuracy to the spectral methods  \cite{BS2005}, which for years were the best choice for such problems. The main advantage of the finite differences is their scalability at the supercomputers. We hope that this paper can be useful for the further development of the realistic  MHD turbulence models with rotation. In spite of the incompressible form of convection's equations, considered above, the substitute of variables ${\bf V}\to \rho{\bf V}$ can be used for transition to the anelastic approximation, suitable to the  dynamo in the compressible medium.


\begin{thebibliography}{100}

\bibitem{Roberts} 
 {Roberts}, P.-H.  
({1968}), 
{On the thermal instability of a rotating-fluid sphere containing heat sources}, 
{\it {Phil.~Trans.~R.~Soc}, {A263}}, 
{93}--117.  

\bibitem{AL1972} 
{Arzimovich}, L., A., S.Yu.~Lukianov 
({1972}), 
{\it {Motion of the charged particles in the electromagnetic fields. In Russian}}, 
Nauka.

\bibitem{Bandaru2016} 
{Bandaru}, V.,      T.~Boeck,  D.~Krasnov, J.~Schumacher 
({2016}), 
{A hybrid finite difference–boundary element procedure for 
the simulation of turbulent MHD duct flow at finite magnetic 
Reynolds number}, 
{\it {J. Comp. Phys.}, {304}}, 
{320}--339.

\bibitem{BS2005} 
{Brandenburg}, A.,  K.~ Subramanian 
({2005}), 
{Astrophysical magnetic fields and nonlinear dynamo}, 
{\it {Phys. Rep.}, {417}}, 
{1}--209.

\bibitem{B2003} 
{Buffett}, B.  
({2003}), 
{A comparison of subgrid-scale models for large-eddy simulations of convection in the Earth's core}, 
{\it {Geophys. J. Int.}, {153}}, 
{\bf 3},      
{753}--765.

\bibitem{Busse} 
{Busse}, F.-H.  
({1970}), 
{Thermal instabilities in rapidly rotating systems}, 
{\it {Fluid Mech.}, {44}}, 
{441}--460.

\bibitem{B2002} 
{Busse}, F.H.  
({2002}), 
{Convective flows in rapidly rotating spheres and their dynamo action}, 
{\it {Phys. Fluids}, {14}}, 
{1301}--1314.

\bibitem{CHZ1988} 
{Canuto}, C.,   M.~Y.~Hussini, Q.A.~Zang 
({1988}), 
{\it {Spectral Methods in Fluids Dynamics}}, 
Springer-Verlag.

\bibitem{CEW2003} 
{Cattaneo}, F.,  T.~Emonet, N.~Weiss 
({2003}), 
{On the interaction between convection and magnetic fields}, 
{\it {Astrophys. J.}, {588}}, 
{\bf 2},      
{1183}--1198.

\bibitem{GRPEPI95} 
{Glatzmaier}, G., P.H.~Roberts 
({1995}), 
{A three-dimensional convective dynamo solution with rotating 
and finitely conducting inner core and mantle}, 
{\it {Phys. Earth Planet. Int.}, {91}}, 
{63}--75.

\bibitem{Jones} 
{Jones}, C. A.  
({2000}), 
{Convection-driven geodynamo models}, 
{\it {Phil. Trans.~R.~Soc.~London}, {A358}}, 
{873}--897.

\bibitem{JR2000} 
{Jones}, C. A., P. H.~Roberts 
({2000}), 
{Convection-driven dynamos in a rotating plane layer}, 
{\it {J. Fluid Mech.}, {404}}, 
{311}--343.

\bibitem{P2012} 
{Pedlosky}, J.  
({2012}), 
{\it {Geophysical fluid dynamics}}, 
Springer-Verlag.

\bibitem{R2017} 
{Reshetnyak}, M.  
({2017}), 
{The anisotropy of hydrodynamical and current helicity}, 
{\it {Astronomy Reports}, {61}}, 
{\bf 9},    
{783}--790.

\bibitem{RP2016} 
{Reshetnyak}, M., V.~Pavlov 
({2016}), 
{Evolution of the Dipole Geomagnetic Field. Observations and Models}, 
{\it {Geomagnetism and Aeronomy}, {56}}, 
{\bf 1},    
{110}--124.

\bibitem{RHK2013} 
{R{\"u}diger}, G.,   R.~Hollerbach, L.L.~Kitchatinov 
({2013}), 
{\it {Magnetic Processes in Astrophysics: Theory, Simulations, Experiments}}, 
Wiley-VCH.


\end{thebibliography}
\end{document}